\documentclass{aa}  

\usepackage{graphicx}
\usepackage{txfonts}
\defcitealias{KK19}{KK19}
\defcitealias{Hasegawa93}{HH93}
\begin{document}

   \title{Evaporative cooling of icy interstellar grains}
   \subtitle{I. Basic characterization}
   \author{Juris Kalv\=ans
          \and
          Juris Roberts Kalnin
          }
   \institute{
Engineering Research Institute "Ventspils International Radio Astronomy Center" of Ventspils University of Applied Sciences,\\
In$\check{z}$enieru 101, Ventspils, LV-3601, Latvia\\
\email{juris.kalvans@venta.lv}
             }

   \date{Received August 7, 2019; accepted December 5, 2019}

  \abstract
   {While radiative cooling of interstellar grains is a well-known process, little detail is known about the cooling of grains with an icy mantle that contains volatile adsorbed molecules.}
   {We explore basic details for the cooling process of an icy grain with properties relevant to dark interstellar clouds.}
   {Grain cooling was described with the help of a numerical code considering a grain with an icy mantle that is structured in monolayers and containing several volatile species in proportions consistent with interstellar ice. Evaporation was treated as first-order decay. Diffusion and subsequent thermal desorption of bulk-ice species was included. Temperature decrease from initial temperatures of 100, 90, 80, 70, 60, 50, 40, 30, and 20\,K was studied, and we also followed the composition of ice and evaporated matter.}
   {We find that grain cooling occurs by partially successive and partially overlapping evaporation of different species. The most volatile molecules (such as N$_2$) first evaporate at the greatest rate and are most rapidly depleted from the outer ice monolayers. The most important coolant is CO, but evaporation of more refractory species, such as CH$_4$ and even CO$_2$ , is possible when the former volatiles are not available. Cooling of high-temperature grains takes longer because volatile molecules are depleted faster and the grain has to switch to slow radiative cooling at a higher temperature. For grain temperatures above 40\,K, most of the thermal energy is carried away by evaporation. Evaporation of the nonpolar volatile species induces a complete change of the ice surface, as the refractory polar molecules (H$_2$O) are left behind.}
   {The effectiveness of thermal desorption from heated icy grains (e.g., the yield of cosmic-ray-induced desorption) is primarily controlled by the thermal energy content of the grain and the number and availability of volatile molecules.}

   \maketitle

\section{Introduction}
\label{intrd}

The typical grain temperature in cold and dark molecular cloud cores is about 10\,K, which is primarily maintained by the interstellar radiation field \citep[ISRF;][]{Hulst46}. Several processes may be able to temporally heat the grains to substantially higher temperatures of some tens of K. These include the ultraviolet photons of the ISRF \citep[for small grains,][]{Duley73}, grain collisions \citep[e.g.,][]{Oort46,Wickra65}, and, notably, heating of medium and large grains by cosmic rays \citep[CRs;][]{Watson72,Leger85}.

Volatile species adsorbed on interstellar grain surfaces undergo evaporation when the grain temperature exceeds about 30\,K \citep[hereafter KK19]{KK19}. The severing of the physisorption bonds of evaporating molecules robs the grain of thermal energy. This is the process of evaporative cooling, which can be much faster than radiative cooling. Evaporation is possible for temporally heated grains in regions where volatile interstellar ice is present on grain surfaces. This requires a sufficiently low ambient grain temperature and protection from the ISRF, which induces nonthermal desorption. Interstellar cloud cores are an example of such regions.

The most important aspect considered in astrochemistry in which evaporative cooling plays a role is CR-induced whole-grain heating. The grain cools down through evaporation of molecules that escape into the gas phase. This CR-induced desorption (CRD) helps maintain a continuous presence of heavy species in the gas even in cold and dark molecular cloud cores \citep[hereafter HH93]{Hasegawa93}.

Evaporation of solid grains in vacuum has been considered in the literature only to a limited extent. \citet{Watson72} considered grain heating by CRs, followed by evaporation. These authors estimated that evaporation takes about 10\,s and that about 10\,\% of the grain energy is carried away by the evaporating molecules. We were unable to find other published research focusing on this specific topic, although evaporation of liquid droplets in the atmosphere is rather well covered, in connection with aeronautical or weather-related research \citep[e.g.,][]{Hardy47,Du13}. The lack of this knowledge is illustrated by the uncertainties and unknown parameters encountered by \citet{Herbst06}, \citet{Iqbal18}, and \citet{Zhao18} in their studies of CRD.

In astrophysics, \citetalias{Hasegawa93} employed evaporative cooling in their description of CRD. These authors considered a 0.1$\mu$m grain, heated by an impact of a CR iron nucleus to a temperature of 70\,K. They estimated the grain cooling timescale to be similar to the evaporation time of CO ($\sim10^{-5}$\,s), an abundant icy species in the interstellar medium \citep[ISM; e.g.,][]{Lacy84} that is also highly volatile. \citetalias{KK19} considered CR-heated grains with a variety of temperatures. For the cooling of grains, this study employed the same general approach as \citetalias{Hasegawa93}, with the addition of radiative cooling. In this study, radiative cooling was important in cooling of grains with temperatures of 40\,K and below.

Several aspects make grain cooling a much more complex process than considered by \citeauthor{Hasegawa93} and \citeauthor{KK19}. First, the result analysis of the latter study revealed that icy molecules (e.g., N$_2$) other than CO may play an important role in the cooling.

To contribute to evaporative cooling, grain energy loss from thermal desorption for must be higher than or comparable to energy loss from emission of photons. This requirement can be fulfilled by a number of icy species that are potentially abundant on grain surfaces:  N$_2$, O$_2$, CO, CH$_4$, and even CO$_2$. The evaporation of the more refractory species becomes possible only if the numbers of more volatile molecules are insufficient for grain cooling. In a case with high initial temperature $T_0$ (and therefore high thermal energy $E$ of the grain) for a medium or large grain covered with a complex ice layer, this may induce a step-like evaporation sequence: as the volatiles are depleted, cooling timescales increase and refractory molecules become increasingly prone to thermal desorption until radiative cooling ends the evaporation cascade.

A second aspect that complicates the cooling through evaporation is that all surface molecules are not immediately available for evaporation. Grain mantles consist of multiple layers \citep[e.g.,][]{Garrod13b} containing rather refractory molecules, such as water and ammonia \citep{Merrill76}, that do not evaporate under typical conditions of grain heating. The evaporation rate is therefore affected by the successive revelation of ice monolayers (MLs) below the initial surface. During evaporation, the proportion of surface refractory species grows as the volatile molecules escape. The situation is made even more complex with the ability of subsurface species to become mobile at elevated temperatures \citep{Oberg09}, and to diffuse out to the surface before they evaporate.

We aim to describe the physical process of evaporative cooling of a grain covered with an interstellar ice mantle and clarify if the existing approximation of this cooling needs to be improved. For this purpose, calculations were performed considering all of the above issues (Section~\ref{clcl}). In addition to satisfying curiosity on how the evaporation unfolds (Section~\ref{rslt}), we pay particular attention to implications on the CRD process (Section~\ref{cncl}). Because basic phenomena had to be investigated, only a single variety of grains was considered.

\section{Calculations}
\label{clcl}

\subsection{Evaporation of icy molecules}
\label{clc-evap}

We consider a spherical grain covered by an icy mantle with constant thickness. The mantle consists of ice molecules arranged in MLs. MLs near the icy surface encompass a larger sphere and thus have a larger area. Grain cooling was assumed to start at point when the grain has an elevated temperature $T_0$, achieved by some event just prior to the simulation.

\subsubsection{Basic equations}
\label{clc-eq}

For a chemical species undergoing desorption from a limited surface, $N$ is the initial amount of molecules, which changes to $N_{\Delta t}$ after a time $\Delta t$. The number of evaporated molecules is
   \begin{equation}
   \label{bas1}
N_{\rm evap} = N - N_{\Delta t} \,.
   \end{equation}

For a single time-step $\Delta t$ for a grain with temperature $T$, $N_{\Delta t}$ for species undergoing thermal desorption was calculated as for a first-order decay,
   \begin{equation}
   \label{bas2}
         N_{\Delta t} = N \times {\rm exp}(-\Delta t/t_{\rm evap}) \,,
   \end{equation}
where $t_{\rm evap}$ is the evaporation timescale,
   \begin{equation}
   \label{bas3}
         t_{\rm evap} = \frac{{\rm exp}(E_D/k_BT)}{\nu_0} \,.
   \end{equation}
Here, $E_D$ is the desorption energy of the volatile species under consideration, $k_B$ is the Boltzmann constant, and $T$ is the grain temperature. $\nu_0$ is the characteristic vibration frequency of molecules of species $j$,
   \begin{equation}
   \label{bas4}
         \nu_{0,j} = \sqrt{ \frac{2 n_S E_D}{\pi^2 m} } \,,
   \end{equation}
where $m$ is the molecule mass and $n_S$ is the number density of adsorption sites \citep[cm$^{-2}$;][]{Hasegawa92}. For the purposes of this model, we calculated the $n_S$ and ice thickness assuming that the average molecule occupies a volume equal to a cube with edge length $3.5\times10^{-8}$\,cm. This corresponds to a watery ice that includes tens of percent other molecules with higher molecular masses. The average density is similar to 1\,g\,cm$^{-3}$, in line with the adopted composition of the ice (Section~\ref{icprop}).

The evaporating molecules rob the grain of energy, prompting recalculation of $T$ and initiation of the next calculation step (Section~\ref{clc-t}). The total number of molecules that is desorbed during evaporative cooling from an initial temperature $T_0$ to the final temperature $T_2$ is
   \begin{equation}
   \label{bas5}
      N_{\rm ev.tot} = \int_{T_0}^{T_2} N_{\rm evap}(T,\Delta t) {\rm d}T\,.
   \end{equation}

\subsubsection{Evaporation from the surface}
\label{clc-rig}

The icy mantles on interstellar grains consist of a variety of species, some of which are volatile and some of which are refractory for the temperatures and timescales of interest. The molecules were arranged in a series of MLs. For dark cloud cores with a complete freeze-out, grains with a typical size of 0.1\,$\mu$m carry about 100\,MLs of ice.

Below we outline the sequence for calculating the number of molecules $j$ that evaporate from the $n$-th layer during the $m$-th time step. The ice contains several volatile evaporating species ($i$ to $k$, including $j$).

As a result of desorption, deeper ice layers become increasingly exposed (become part of the surface). For the $m$-th step, the part $X_{n,m}$ of all molecules in layer $n$ that have become exposed to evaporation depends on the amount of molecules evaporated from the layer above in the previous step,
   \begin{equation}
   \label{rig1}
   X_{n,m} = \frac{\sum_i^k N_{{\rm evap},j,n-1,m-1}}{N_{n-1}} \,.
   \end{equation}
Here,  $N_{n-1}$ is the number of adsorption sites in the ice ML above the $n$-th layer and $\sum_i^k N_{{\rm evap},j,n-1,m-1}$ is the total number of molecules of all evaporated species (i.e., vacated adsorption sites) in the previous time step $m-1$. There are a few special cases for $X_{n,m}$. For the first layer (outer surface ML), $X_{1,m}$ is always unity. In the first time step, when evaporation has not yet started and all MLs are complete, $X_{n,1}$ is zero for all MLs, except for the first ML.

The number of molecules $j$ in a layer $n$ exposed to evaporation during the current time step $m$ was calculated by modifying the respective value from the previous time step $N_{X,j,n,m-1}$ by adding the number of exposed molecules and subtracting evaporated molecules from the previous time step $m-1$,
   \begin{equation}
   \label{rig2}
   N_{X,j,n,m} = N_{X,j,n,m-1} + X_{n,m} N_{j,n} - N_{{\rm evap},j,n,m-1} \,.
   \end{equation}
$N_{j,n}$ is the initial number of $j$ molecules in the $n$-th layer, equal to the number of adsorption sites in that layer; this is a constant. For the first time step, the number of evaporated molecules is zero. By paralleling Equations (\ref{bas1}) and (\ref{bas2}), the number of evaporated molecules from $N_{X,j,n,m}$ during the current step $m$ can be calculated,
   \begin{equation}
   \label{rig3}
         N_{{\rm evap},j,n,m} = N_{X,j,n,m} -   N_{X,j,n,m} \times {\rm exp}(-t_m/t_{{\rm evap},j}) \,,
   \end{equation}
where $t_m$ is the length of the $m$-th time step ($\Delta t$ in seconds) and $t_{\rm evap,j}$ is the characteristic evaporation timescale for $j$ at the current temperature in the $m$-th step, following Equation~(\ref{bas3}). At this point, we start the next time step by returning to Equation~(\ref{rig1}),
   \begin{equation}
   \label{rig4}
   X_{n,m+1} = \frac{\sum_i^k N_{{\rm evap},j,n-1,m}}{N_{n-1}} \,.
   \end{equation}

\subsubsection{Diffusive evaporation}
\label{clc-dff}

Experiments with interstellar ice analogs have shown that volatile molecules in bulk ice can escape from the H$_2$O matrix at elevated temperatures \citep[e.g.,][]{Collings04}. This means that the molecules undergo subsurface diffusion. This makes the calculation of evaporation rate more complex because buried volatile molecules can now also undergo evaporation.

In this research, we are interested only in a molecular diffusion that eventually results in thermal desorption, that is, diffusion from bulk ice MLs to the surface, followed by evaporation. The exchange of molecules between bulk-ice layers was ignored for the sake of simplicity and to allow us to focus on processes that result in evaporation. The diffusion rate was calculated according to \citetalias{KK19}, Equation~(1). The time of diffusion to the surface layer of a molecule ($n-1$) MLs below the surface is thus
   \begin{equation}
   \label{dff1}
   t_{{\rm diff},j,n} = t_{{\rm hop},j} \times 6 (n-1)^2 \,,
   \end{equation}
where the hopping time between the MLs was calculated in analogy with the evaporation time of Equation~(\ref{bas3}),
   \begin{equation}
   \label{dff2}
         t_{{\rm hop},j} = \frac{{\rm exp}(E_{b,j,n}/k_BT)}{\nu_{1,j}} \,.
   \end{equation}
Here, $E_{b,j,n}$ is the binding energy for species $j$ in the $n$-th layer, discussed in Section~\ref{icprop}. This barrier has to be overcome by bulk-ice molecules to change their location to an adjacent absorption site (molecule-sized cell), discussed in Section~\ref{icprop}. $\nu_1$ was calculated similarly to $\nu_0$ in Equation~(\ref{bas4}), but by assuming that bulk-ice molecules have desorption energies equal to $2E_D$, following \citet{Garrod13a}.

In the diffusive mantle, surface evaporation and evaporation of diffused bulk-ice molecules was treated separately. Both mechanisms affect each other's rate by changing the number of molecules in ice MLs. Surface evaporation depends on $N_{X,j,n,m}$, the number of exposed molecules $j$ in layer $n$ at time step $m$ (Equation~(\ref{rig3})), while diffusive evaporation occurs for molecules that are still covered by other MLs above. The number of these molecules is abbreviated $N_{C,j,n,m}$. Again, by paralleling Equation (\ref{bas1}), the number of molecules $j$ that evaporated through diffusion is
   \begin{equation}
   \label{dff3}
         N_{{\rm ev-diff},j,n,m} = N_{C,j,n,m} -        N_{C,j,n,m} \times {\rm exp}\left(\frac{-t_m}{t_{{\rm evap},j}+t_{{\rm diff},j,n}}\right) \,.
   \end{equation}
The number of the covered molecules is reduced in each step according to
   \begin{equation}
   \label{dff4}
   N_{C,j,n,m} = N_{C,j,n,m-1} - N_{{\rm ev-diff},j,n,m} - X_{n,m} (N_{j,n}-N_{{\rm ev-diff},j,n}) \,,
   \end{equation}
where the number of $j$ molecules diffused out so far (until the $m$-th step) is
   \begin{equation}
   \label{dff5}
   N_{{\rm ev-diff},j,n} = \sum_1^m N_{{\rm ev-diff},j,n,m} \,.
   \end{equation}
The quantity $X_{n,m}(N_{j,n}-N_{{\rm ev-diff},j,n})$ is the number of $j$ molecules exposed to the outer surface during the current time step $m$ in the $n$-th ML. The number of molecules exposed to the surface for possible evaporation changes according to
   \begin{equation}
   \label{dff6}
   N_{X,j,n,m} = N_{X,j,n,m-1} + X_{n,m} N_{C,j,n,m} - N_{{\rm evap},j,n,m-1} \,.
   \end{equation}
In the case of a diffusive mantle, evaporation from the surface and the bulk-ice occurs concurrently, and the above equation replaces Equation~(\ref{rig2}), which does no consider diffusion. The number of molecules that evaporated from the surface $N_{{\rm evap},j,n,m}$ and the proportion of the $n$-th ML that is exposed to the surface $X_{n,m}$ are still calculated with Equations (\ref{rig3}) and (\ref{rig4}).

The total number of molecules $j$ that evaporated during the $m$-th time step is equal to
   \begin{equation}
   \label{dff7}
N_{{\rm ev.tot,}j,m} = \sum_1^{n_{\rm tot}} (N_{{\rm evap},j,n,m} + N_{{\rm ev-diff},j,n,m})\,,
   \end{equation}
where $n_{\rm tot}$ is the number of ice MLs. By adding $N_{{\rm ev.tot,}j,m}$ step by step, we obtained the total number $N_{{\rm ev.tot,}j}$ of molecules $j$ that evaporated throughout the entire simulation. This quantity corresponds to $N_{\rm ev.tot}$ in Equation~(\ref{bas5}). The number of icy molecules remaining in the MLs during a given step is the aggregate of exposed and covered $j$ populations,
   \begin{equation}
   \label{dff8}
N_{j,n,m} = N_{X,j,n,m} + N_{C,j,n,m}\,.
   \end{equation}

\subsection{Adopted properties of ice}
\label{icprop}
%
   \begin{figure}
    \vspace{-2cm}
    \hspace{-2cm}
    \includegraphics[width=24cm]{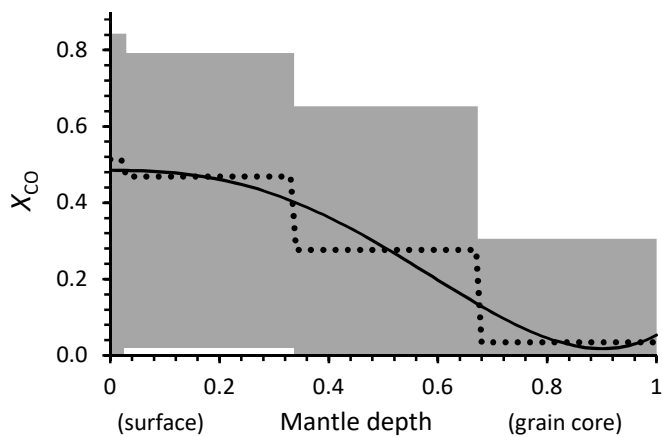}
    \vspace{-26cm}
   \caption{Example of source data for the CO molecule for obtaining ice molecule abundances as a function of ice depth. The $y$-axis indicates the part of an ice layer that consists of CO, while the $x$-axis gives the mantle depth from the surface in decimal parts (in Figure~\ref{att-ini} it is multiplied by 100, the number of MLs). The dotted line is the average relative abundance of CO in ice, and the shaded area represents the interval of possible values, while the solid line is the CO abundance function used in the model, Equation~(\ref{ic3}); this is a trend line of the average values.}
              \label{att-poly}
    \end{figure}
%
   \begin{figure*}
   \centering
    \includegraphics[width=17cm]{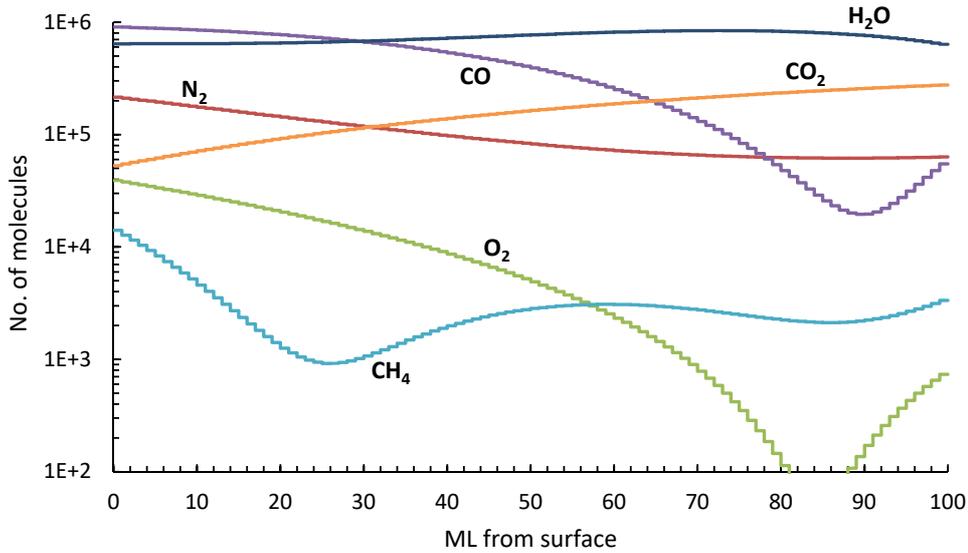}
    \vspace{-15cm}
   \caption{Initial ice composition per ML, as used in the model.}
              \label{att-ini}
    \end{figure*}
%
\begin{table}
\caption{Desorption energies and the overall abundance of the icy species relative to water.}
\label{tab-ini}
\centering
\begin{tabular}{l c r}
\hline\hline
 & & Abundance, \\
Species & $E_D$, K & \% of water \\
\hline
N$_2$ & 1000 & 13.9 \\
O$_2$ & 1000 & 1.4 \\
CO & 1150 & 56.7 \\
CH$_4$ & 1300 & 0.4 \\
CO$_2$ & 2580 & 22.2 \\
H$_2$O & 5700 & 100.0 \\
\hline
\end{tabular}
\end{table}
We considered a mantle consisting of compact nonporous ice, which is expected to be similar to the case of interstellar ice \citep{Jenniskens95}. When the grain is heated, evaporation arises, including diffusive evaporation from below the surface. We assumed that the mobility of molecules that ensures diffusion and desorption also allows the remaining ice molecules to occupy energetically favorable adsorption and absorption sites. This keeps the ice compact, while the grain always remains spherical.

To properly model processes in interstellar ice, we need a reliable estimate on the composition of the ice. Our aim is to model cooling for grains that are relevant to the cores of dark molecular clouds. Observations of interstellar ice in these cores indicate polar and nonpolar ice \citep{Tielens91} and give information on the proportions of different icy species \citep[e.g., see the references in Table~A1 of][]{K18mn}. However, these data are insufficient for deriving the composition of a typical icy mantle with a resolution of one ML.

Detailed information on the composition of different ice MLs can fortunately be derived from astrochemical models. For this task, we employed the results of our previous modeling studies \citep[\citetalias{KK19}]{K15apj1,K15apj2,K18iau,K18mn,K17} considering the chemistry in starless and prestellar cores. These models follow the evolution of a cloud core that undergoes delayed gravitational collapse up to a certain maximum gas density in the range $10^5...10^7$\,cm$^{-3}$. Some simulations continue, either considering the heating of a protostellar envelope or steady-state chemical evolution of a stable starless core. The calculated ice composition for each simulation was sampled at the point when maximum density is reached for the first time. This point typically coincides with the freeze-out of heavy molecules. For simulations extending beyond this point, the chemical abundances of ice were sampled for the second time at integration times 2.0\,Myr for \citet{K15apj2}, 0.1\,Myr for \citet{K17}, and 2.5\,Myr for \citetalias{KK19}. These particular times were chosen so that they ensured that the sampled relative abundances of ice components remained in line with observational constraints and were relevant for cloud cores that do not contain stars.

The average abundances of five major and potentially volatile icy species,  N$_2$, O$_2$, CO, CH$_4$ , and CO$_2$ , were obtained for data from each of the above-mentioned papers. In turn, these values were averaged again, obtaining a single value for each `sublayer'. All these models describe the icy mantles as consisting of four sublayers: the surface, and three bulk-ice layers, with averaged thickness proportions 2:31:33:34, respectively. A trend line of these data was then derived to obtain the species' proportion $X$ in ice as a function of ice depth $n$, measured in MLs, as illustrated in Figure~\ref{att-poly}. Minimum and maximum values of $X_n$ were also obtained to ensure that the trend line did not exceed the limits provided by the calculations in these published studies. The abundance functions were expressed with the help of polynomials:
   \begin{equation}
   \label{ic1}
   X_{{\rm N_2},n} = 0.1335x^2 - 0.1869E-1x + 0.1164 \,,
   \end{equation}
   \begin{equation}
   \label{ic2}
   X_{{\rm O_2},n} = 0.03292x^2 - 0.05287x + 0.02123 \,,
   \end{equation}
   \begin{equation}
   \label{ic3}
   X_{{\rm CO},n} = 1.722x^4 - 2.145x^3 + 0.2029x^2 - 0.2340x + 0.4999 \,,
   \end{equation}
   \begin{equation}
   \label{ic4}
   X_{{\rm CH_4},n} = 0.1340x^4 - 0.3104x^3 + 0.2463x^2 - 0.07486x + 0.008230 \,,
   \end{equation}
   \begin{equation}
   \label{ic5}
   X_{{\rm CO_2},n} = 0.1195x^2 + 0.1176x + 0.02771 \,,
   \end{equation}
   \begin{equation}
   \label{ic6}
   X_{{\rm H_2O},n} = 1 - X_{{\rm N_2},n} - X_{{\rm O_2},n} - X_{{\rm CO},n} - X_{{\rm CH_4},n} - X_{{\rm CO_2},n} \,.
   \end{equation}

Here, $x=n/n_{\rm tot}$, that is, $x$ is depth of the n-th ML expressed as part of the mantle thickness.  Equation~(\ref{ic6}) means that the remainder of the ice was assumed to consist of refractory species, primarily water. Figure~\ref{att-ini} shows the ice composition per ML according to Equations (\ref{ic1}-\ref{ic6}), while Table~\ref{tab-ini} lists the overall relative abundance of the icy species, as well as their desorption energies.

Temperature-programmed desorption experiments with interstellar ice analogs (icy mixtures) show that volatile molecules may escape the water-ice matrix. Some molecules diffuse to the surface and evaporate, and some remain trapped in bulk ice. This desorption stops when the surface becomes saturated with water ice or other refractory species \citep{Fayolle11,Martin14}. Molecules buried deeper in the ice are less likely to escape, but small and nonpolar species with low $E_D$ can escape from deeper layers \citep{Oberg09}.

In calculating the number of molecules that is removed from ice through diffusive evaporation, we employed the following approach on energy barriers. The binding energy $E_{b,n}$ (for Equation~(\ref{dff2})) that has to be overcome for a bulk-ice molecule residing in the second ML (immediately below the surface) for diffusion to the surface was taken to be 0.55 times its desorption energy ($2E_D$, see Equation~(\ref{dff2})), i.e. $E_{b,2}=1.1E_D$. The multiplier 0.55 was adopted from the diffusion-to-desorption energy ratio for surface species, as determined by \citet{Minissale16}. This $E_b$ value was then linearly increased with depth, layer by layer. After an assumed number $c$ of layers, $E_b$ (of any molecule) reaches a maximum of 5700\,K, the $E_D$ of water, meaning that the molecule has become immobile, unless the water-ice lattice itself becomes unstable. Thus, the binding energy of bulk ice for icy species $j$, residing in the $n$-th ML, was calculated as 
   \begin{equation}
   \label{ic7}
   \begin{array}{l}
	E_{b,j,n} = 1.1E_{D,j} \frac{c-(n-2)}{c} + E_c \frac{(n-2)}{c} \\
	E_{b,j,n} \leq E_{D,{\rm H_2O}} \,.
   \end{array}
   \end{equation}
The parameter $E_c$ is a constant, here taken to be equal to the water absorption energy, $2E_{D,{\rm H_2O}}$. $E_b$ was never allowed to be higher than $E_{D,{\rm H_2O}}$ for any species, except for water, whose $E_{b,{\rm H_2O}}$ was taken to be $1.1E_{D,{\rm H_2O}}$ in all bulk-ice MLs. Molecules in the first ML (surface) do not undergo diffusive evaporation, and their binding energy $E_{b,j,1}$ is irrelevant in this model. Volatile species with smaller $E_D$ reach this value in deeper layers and thus are able to diffuse out from greater depths. Higher grain temperature also encourages escape from greater depth. Our implemented approach takes into account that the ice thickness may decrease as a result of evaporation.

\citet{Fayolle11} found diffusion depths of 20.5 and 80\,MLs, respectively, for CO$_2$ and CO. At laboratory timescales and 100\,K temperature, the diffusion can be estimated to be possible with energy barriers up to about 3200\,K (from Equations (\ref{dff1}) and (\ref{dff2}); 100\,K is the maximum grain temperature considered in this study). The value of the parameter $c$ was taken to be 450\,ML; this caused CO and CO$_2$ diffusion to occur from MLs down to approximately the values indicated by \citeauthor{Fayolle11}. Our model therefore describes evaporation after icy molecule diffusion and trapping of volatiles below a layer of refractory molecules, with a resolution of 1\,ML.

As the molecules leave, the size of the grain decreases, as does the surface area $\pi r^2$, where $r$ is the total radius (size of the olivine core $a$ plus ice thickness) of the grain. This was taken into account by matching the total number of icy molecules with the number of adsorption and absorption sites available in an icy mantle of a certain thickness. With this model, the ice thickness was reduced by up to a maximum of 13\,MLs for $T_0=100$\,K. The resolution of the grain size is 1\,ML, that is, no partial layers were considered in the model.

\subsection{Change in grain temperature}
\label{clc-t}

Each molecule $j$ robs the grain of a certain amount of energy $E_{{\rm evap},j}$. For the special case with low $T$, $E_D\gg kT$, and considering a large number of molecules, the energy lost by the grain can be estimated as their $E_D$ plus the average remaining energy of the departing molecules,
   \begin{equation}
   \label{t2}
	E_{{\rm evap},j} = \frac{\int_{E_{D,j}}^{\infty} E_j {\rm exp}(E_j/kT) {\rm d}E_j}
	{\int^{\infty}_{E_{D,j}}{\rm exp}(E_j/kT){\rm d}E_j} = E_{D,j} + kT \,,
   \end{equation}
where $E_j$ is the energy of molecule $j$ residing on grain surface. The result means that the average surplus energy with which the molecule leaves the grain is equal to $kT$, while energy equal to $E_{D,j}$ is consumed in the severing of the physisorption bond and is thus also lost from the grain. The simple and interesting finding of evaporating molecules having energy $kT$ can be interpreted as follows. The average energy of surface molecules is $kT$. At a given time instant, a small proportion of molecules exceed (or receive a kick) and traverse the energy barrier $E_D$ while keeping a tail of surplus energy, which on average still corresponds to the temperature of the grain at the evaporation time of the molecule.

The energy carried away by all molecules $j$ is $N_{{\rm ev.tot},j}E_{{\rm evap},j}$. The total energy carried away by all volatiles $i$ through $j$ to $k$ is
   \begin{equation}
   \label{t6}
	E_{\rm evap} =  \sum^k_i{N_{{\rm evap},j}E_{{\rm evap},j}}\,.
   \end{equation}

Alongside evaporation, radiative cooling takes place. The energy lost through radiative cooling is $E_{\rm rad}$. It was calculated for each time step $m$ by solving Equation (21) of \citet{Cuppen06},
   \begin{equation}
   \label{t7}
	\frac{E_{\rm rad}}{t_m} =  4\pi^2 r^2 \int_{\rm91.2\,nm}^{\rm1\,cm} Q_{\rm em}(\lambda) B_\lambda(T) {\rm d}\lambda \,,
   \end{equation}
where $Q_{\rm em}$ is the emission coefficient and $B$ is the Planck function. Here we did not distinguish the icy mantle from the olivine grain core. While we considered the emission of photons in the wavelength range of 91.2\,nm to 1\,cm, only photons with $\lambda>5\,\mu$m contribute to grain cooling with $T\leq100$\,K. Radiative cooling becomes more effective than surface evaporation of N$_2$ at temperatures of 31\,K and below. This can be compared to the 34\,K threshold obtained with the simpler radiative cooling approach of \citet{Duley73}, which we used in our previous study \citepalias{KK19}.

The thermal energy lost by the grain during the cooling process is
   \begin{equation}
   \label{t8}
	E_{\rm cool} = E_{\rm evap} + E_{\rm rad}\,.
   \end{equation}
The thermal energy changes for the grain were followed by obtaining the $E_{\rm evap}$ and $E_{\rm rad}$ for time steps of length $t_m$ with a fixed current temperature $T$. The loss of the thermal energy $E_{\rm cool}$ of the grain during the $m$-th time step was used to calculate a new temperature according to
   \begin{equation}
   \label{t9}
	T_{m+1} = T_m - E_{\rm cool}/C(T_m) \,.
   \end{equation}
$C(T_m)$ is the heat capacity of the grain at a temperature $T_m$. $C(T_m)$ was calculated separately for the icy mantle (with current thickness $n_{tot,m}$) and the 0.1\,$\mu$m olivine grain core, according to \citet{K16}. 

\subsection{Model resolution and limitations}
\label{inststp}

The length $t_m$ (or $\Delta t$) of the time steps for the simulations was calculated as a function of the simple radiative cooling rate by \citet{Duley73}. At low temperatures, simple proportionality is sufficient. At high temperatures, where evaporation dominates, a steeper function has to be used. Moreover, regardless of $T_0$ (except for $T_0<30$\,K), rapid evaporation occurs directly from the surface during the first steps of each simulation, and the temperature also rapidly falls. This requires shorter $t_m$ at the beginning of each simulation.

The number of molecules that evaporated during the initial stages subsequently affects the cooling at lower temperatures, when evaporation is regulated by the gradual revelation of subsurface layers and diffusion. These different cooling patterns required unique $t_m$ functions for each simulation with differing $T_0$.  Typically, 100 to 10000 molecules of a single species evaporate during a single time step. Only small decreases in the grain temperature ($\Delta T$ similar to or smaller than 0.01\,K) were tolerated. Test calculations with one hundred times shorter time steps changed the resulting cooling times only by slightly more than 0.1\,\%, therefore we considered the accuracy of the simulations to be satisfactory.

The minimum length of $\Delta t$ is limited by the physics included in the model. For timescales shorter than about $10^{-10}$\,s, heat transfer within the grain has to be considered. Additionally, molecules departing from the grain with a thermal velocity cover a distance that is equal to the grain radius within about $10^{-9}$\,s. Thus, models of high-temperature grains undergoing massive evaporation at these timescales also need to consider the pressure of the evaporated gas. Finally, the calculation of the grain heat capacity has an upper application limit of 150\,K.

\section{Results}
\label{rslt}

We implemented this sequence of calculations in a \textsl{Fortran} program that we called ``TCOOL''. Because we did not find previous studies that would have considered evaporation of grains in any detail, this research focuses on explaining the main qualitative aspects related to the evaporation process. Two complementary viewpoints have to be considered: first, the evolution of the grain temperature after a heating event, and second, the chemical consequences resulting from evaporation. The latter aspect considers changes in the composition of the icy mantle and the number of different molecules desorbed, which in interstellar clouds affects the chemistry of the gas phase.

We considered only `standard' grains with an olivine core of size of 0.1\,$\mu$m for consistency and ice mantle thickness of 100\,MLs. These parameters are relevant for grains in dark cloud cores. The ice composition was described with Equations (\ref{ic1}-\ref{ic5}). Numerous combinations for grain size, mantle thickness, and ice composition are possible, and the exploration of these is reserved for a study in the near future. Impacts of heavy CR ions can heat these standard grains to a temperature of about 90\,K \citep{K16}.

Unlike the case of radiative cooling, evaporative cooling occurs in an unique way for different initial temperatures. This is because previous evaporation affects the number of molecules that is prone to evaporation during a given time step with a given temperature. Therefore, it makes sense to explore cooling from different temperatures. Similarly to \citetalias{KK19}, we employed a set of initial temperatures $T_0$, where each value differed by 10 degrees, in the range 20--100\,K (i.e., 100, 90, 80, 70, 60, 50, 40, 30, and 20\,K). The final temperature was always assumed to be $T_2=10$\,K, a standard value for grains in dark clouds. 

\subsection{Evolution of the grain temperature}
\label{res-T}
%
   \begin{figure*}
   \centering
    \includegraphics[width=20cm]{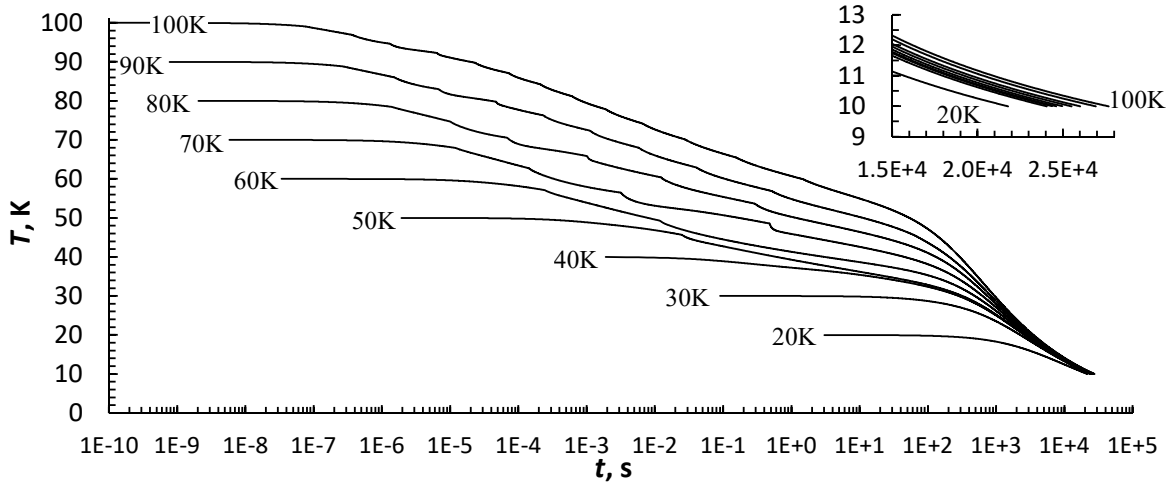}
		\vspace{-20.5cm}
   \caption{Calculated temperature for icy grains with different initial temperatures $T_0$ (K). The inset shows the last few degrees of cooling (before the stop at 10\,K), which takes most of the entire cooling time.}
              \label{att-T}
    \end{figure*}
%
   \begin{figure*}
   \centering
    \includegraphics[width=18cm]{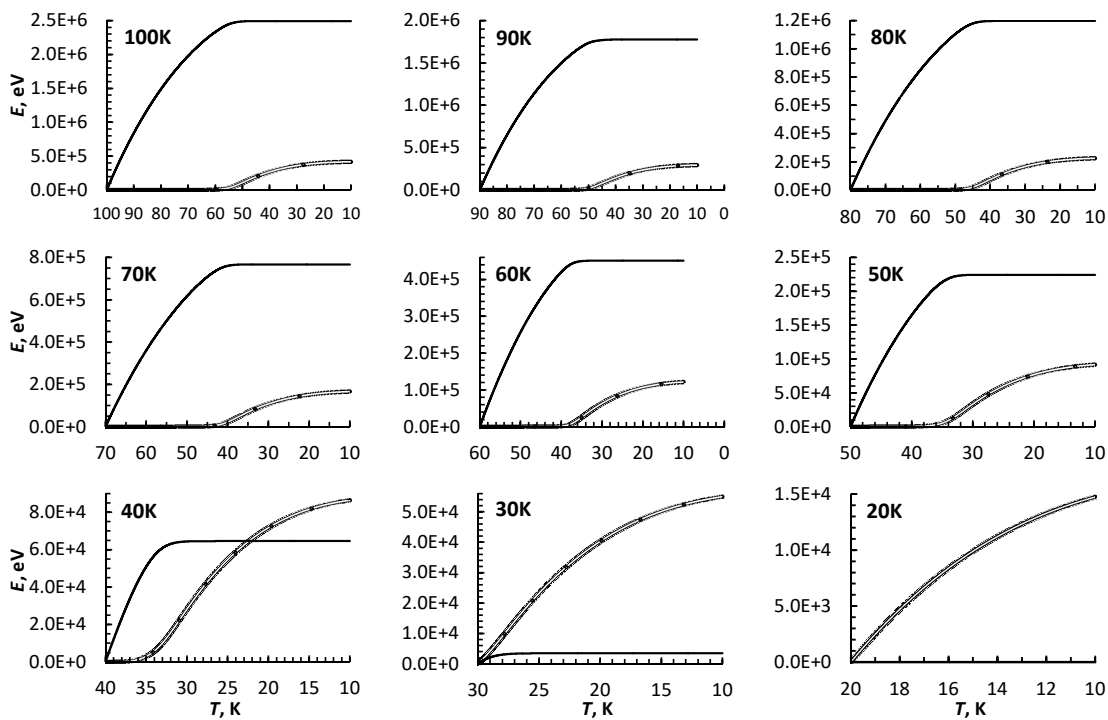}
		\vspace{-14cm}
   \caption{Energy released during cooling from different initial grain temperatures $T_0$. The vertical axis shows the amount of energy that is lost by the grain during cooling, and the horizontal axis shows the evolving grain temperature. The solid line represents the energy lost through evaporation, and the double line plots the radiated energy.}
              \label{att-E}
    \end{figure*}
%
\begin{table*}
\caption{Indicative results on grain cooling from different initial grain temperatures $T_0$ down to the ambient grain temperature of 10\,K.}
\label{tab-E}
\centering
\begin{tabular}{r l r l r c}
\hline\hline
$T_0$ & $E_{\rm cool}$, MeV & \% $E_{\rm rad}$ & $T_{\rm switch}$, K & $t_{\rm switch}$, s  & Final MLs \\
\hline
100\,K & 2.91 & 14.4 & 54.0 & 14.7 & 87 \\
90\,K & 2.07 & 14.2 & 48.8 & 19.1 & 90 \\
80\,K & 1.42 & 15.8 & 44.8 & 22.9 & 94 \\
70\,K & 0.933 & 17.8 & 40.9 & 28.1 & 96 \\
60\,K & 0.572 & 21.2 & 37.3 & 29.5 & 98 \\
50\,K & 0.316 & 29.0 & 34.1 & 46.1 & 99 \\
40\,K & 0.151 & 57.2 & 33.6 & 45.5 & 100 \\
30\,K & 0.0584 & 94.1 & 30.0\tablefootmark{a} & 0.0 & 100 \\
20\,K & 0.0147 & 100.0 & 20.0\tablefootmark{a} & 0.0 & 100 \\
\hline
\end{tabular}
\tablefoottext{a}{Cooling dominated by radiation from the start.}
\tablefoot{$E_{\rm cool}$ is the total released thermal energy of the grain, with the proportion of energy radiated away as photons $E_{\rm rad}$ given as the percentage of $E_{\rm cool}$. $T_{\rm switch}$ and $t_{\rm switch}$ are the temperature and time for switching from evaporative to radiative cooling. The last column gives the final ice thickness after the grain has cooled.}
\end{table*}
Figure~\ref{att-T} shows the evolution of the grain temperature during cooling. The wavy structure for some of the curves is an artifact that arises because the model only considers full MLs of icy molecules (no partially filled layers). Each time the number of MLs is reduced by one (due to evaporation), some molecules in the bulk-ice layers suddenly become more susceptible to diffusion and subsequent evaporation, according to Equations (\ref{dff1}) and (\ref{ic7}). This results in irregularities in the momentary cooling rate.

Comparing the temperature curves for different initial grain temperatures $T_0$, we find that cooling from higher $T_0$ takes much longer (see the inset of Figure~\ref{att-T}). This can be  attributed to any objects undergoing evaporative cooling with a limited pool of volatiles. In the modeled case of icy interstellar grains, any easily evaporating molecules are rapidly removed at high temperatures. This means that a grain cooling from $T_0=100$\,K will have fewer available volatile molecules at 70\,K than a grain that has been heated to just $T_0=70$\,K. Instead, the continued cooling of the grain with $T_0=100$\,K has to rely on molecules that have higher energy barriers and longer evaporation timescales. These are volatiles that reside deeper below the icy surface of the grain, or species that otherwise could be regarded as refractory, such as CO$_2$.

The cooling curve is regulated by evaporation until the grain reaches a point when photons carry away more energy than molecules. This transition occurs gradually because the surface volatiles have long been depleted, and for volatiles in bulk ice, it takes increasingly longer to diffuse to the surface at the current temperature $T_m$ because molecules in shallower layers diffuse out first, while $T_m$ continues to decrease. The transition or switching from a cooling process controlled by evaporation to radiation-dominated cooling can be characterized by the temperature $T_{\rm switch}$ and time $t_{\rm switch}$, when $E_{\rm rad}$ for the current step becomes greater than $E_{\rm evap}$. $T_{\rm switch}$ and $t_{\rm switch}$ are unique for each $T_0$ regime, and are listed in Table~\ref{tab-E}.

The transition to radiation is significant because radiative cooling takes longer and the grain spends more time in an elevated temperature.  Figure~\ref{att-E} shows that $E_{\rm rad}$ starts to grow significantly only shortly before $T_{\rm switch}$. At 30\,K or below, the timescale for even direct surface evaporation of $N_2$ is longer than the radiative cooling timescale, and grain cooling is dominated by radiation.

\subsection{Evaporated molecules}
\label{res-evap}
%
   \begin{figure*}
   \centering
    \includegraphics[width=20cm]{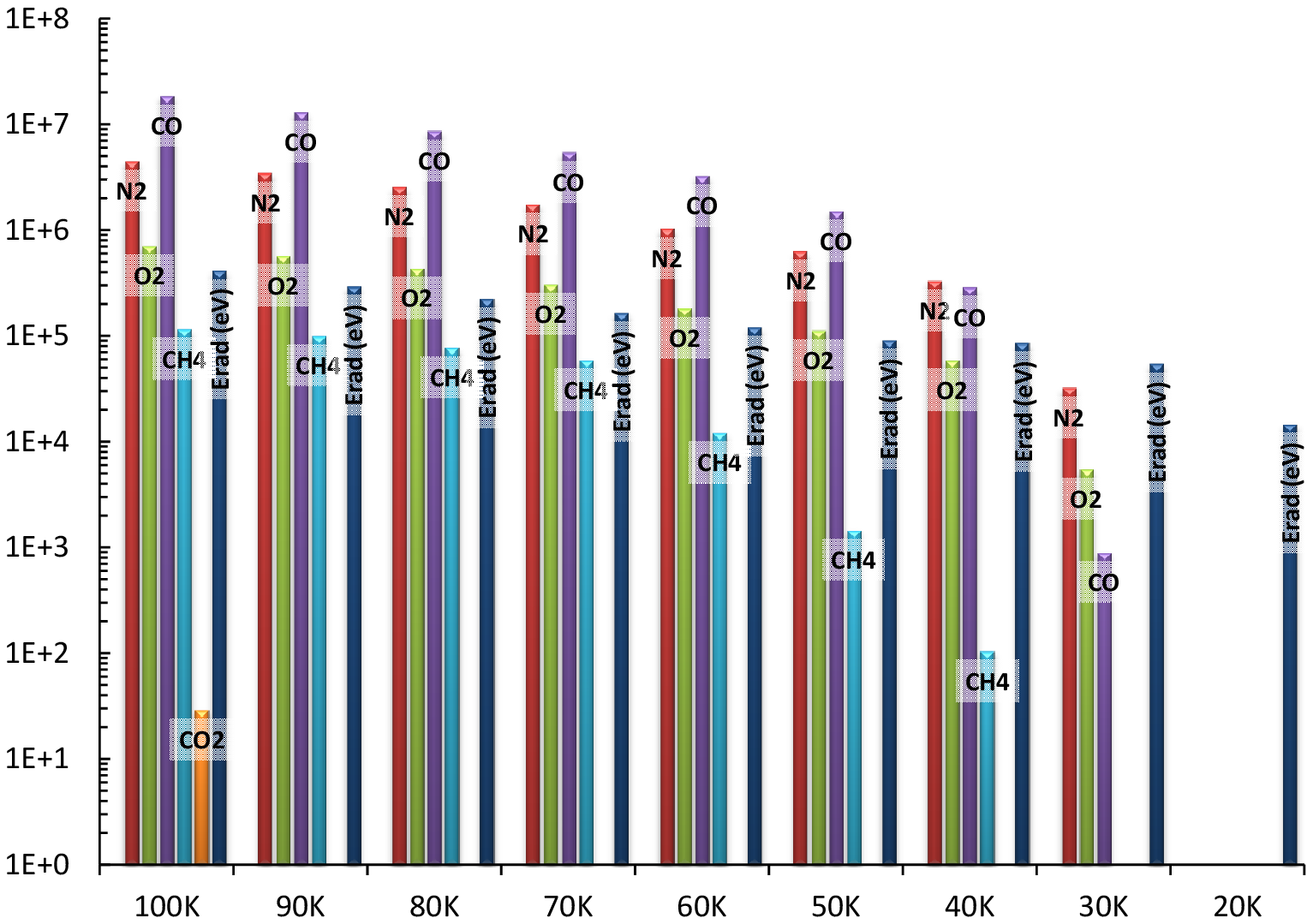}
		\vspace{-16cm}
   \caption{Comparison of the total numbers of evaporated molecules in simulations with different initial grain temperatures $T_0$.}
              \label{att-tot}
    \end{figure*}
%
   \begin{figure*}
   \centering
    \includegraphics{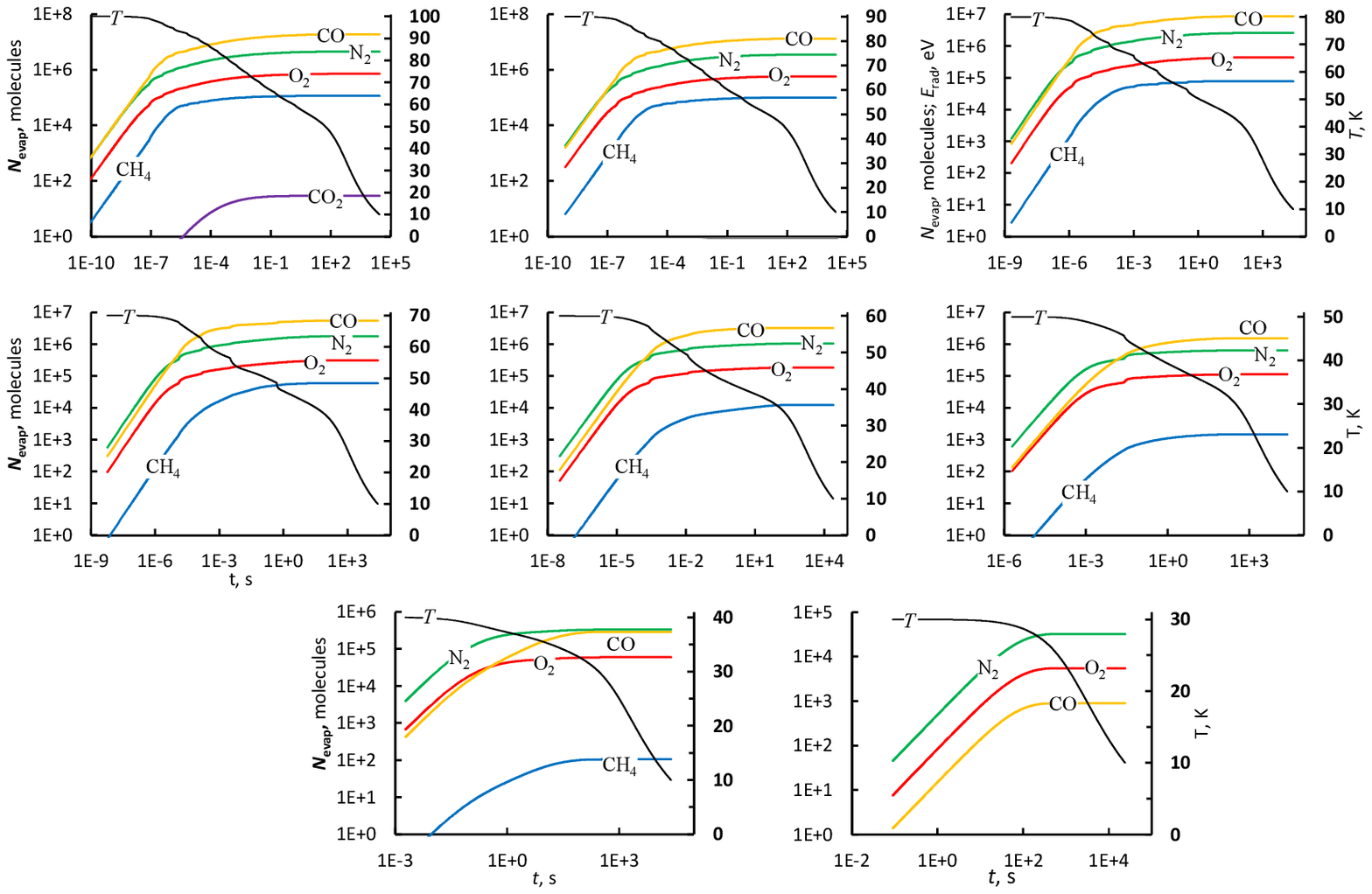}
		\vspace{-17cm}
   \caption{Evolution of the numbers of icy molecules that evaporated during cooling from different initial grain temperatures $T_0$. The temperature curve is shown for context. $T_0$ can be read as the maximum value of the secondary vertical axis on the right-hand side of each plot. The $T_0=20$\,K regime has no evaporated molecules.}
              \label{att-evap}
    \end{figure*}
Figure~\ref{att-tot} shows that an appreciable number of volatiles are evaporated by all $T_0$ regimes, except for that with $T_0=20$\,K. Even the $T_0=40$\,K and $T_0=30$\,K regimes release tens of thousands of molecules, even though the majority of grain energy in these cases is lost through radiation.

Only the regime with $T_0=100$\,K is able to evaporate any number of CO$_2$ molecules. Only tens of these molecules are evaporated, largely because CO$_2$ primarily resides deep below the surface of ice in our model (Figure~\ref{att-ini}). Side calculations with the model revealed that this situation changes for grains with only few volatiles and available surface CO$_2$, where the latter can serve as a major cooling agent.

Table~\ref{tab-E} shows that the majority of thermal energy for high-$T_0$ grains is lost through evaporation. The total number of evaporated molecules therefore primarily depends on the heat content of a grain, not on its cooling time, as was assumed by \citetalias{Hasegawa93}. The other important parameter is the number of accessible molecules. Grains with higher $T_0$ release more molecules \textit{\textup{and}} lose more energy through radiation. Table~\ref{tab-E} shows that the proportion of $E_{\rm rad}$ grows with $T_0$ until 90\,, when there is a sufficient number of volatile molecules for evaporation. The proportion of $E_{\rm rad}$ for the $T_0=100$\,K regime is higher than that for $T_0=90$\,K because the lack of volatiles has taken effect.

Figure~\ref{att-evap} shows details of the evaporation progress. The number of evaporated molecules grows most rapidly at the beginning of simulations, and the evaporation rate decreases exponentially until evaporation becomes insignificant at approximately $t_{\rm switch}$. CO and N$_2$ are the primary coolants, while O$_2$ and CH$_4$ are of secondary importance. Because methane has the highest $E_D$, its evaporation numbers vary more strongly with temperature. Except for the $T_0=100$\,K simulation, in all cases N$_2$ evaporation is fastest; CO takes over when N$_2$ has been depleted.

\subsection{Ice composition}
\label{rslt-ic}
%
   \begin{figure*}
   \centering
    \includegraphics{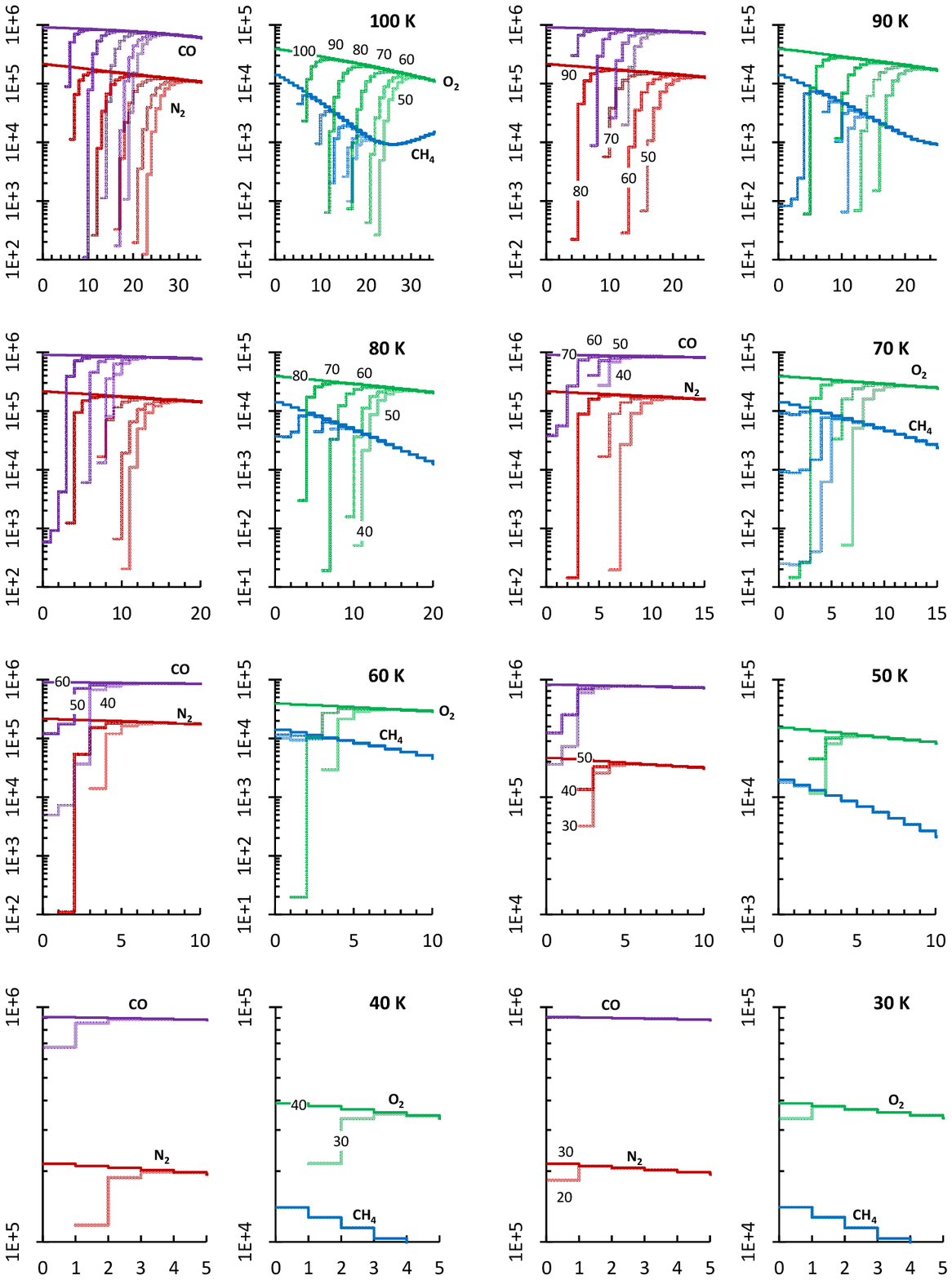}
		\vspace{-8cm}
   \caption{Composition of ice for simulations with different $T_0$, given at several temperatures (multiples of 10\,K) that occur during cooling. Low temperatures are not shown because the ice composition changes little after $T_{\rm switch}$. The results for each $T_0$ are depicted in a pair of plots, with the respective temperatures indicated for only one of the species, but the temperature is also relevant for other species for the given $T_0$. For clarity, the respective curves for each species are indicated only in every other plot. Water and carbon dioxide are not included because their numbers remain unchanged. The curve marked with the value of $T_0$ is always the initial ice composition, fully depicted in Figure~\ref{att-ini}.}
              \label{att-lay}
    \end{figure*}
Figure~\ref{att-lay} shows the depletion of icy volatiles at several points for each simulation. We note that molecules remain associated with their parent ML for the entire duration of the simulation, but the properties of molecule populations in each ML (e.g., the number of the above MLs in Equation~(\ref{dff1})) were calculated correctly by evaluating the actual ice thickness in each integration step. Table~\ref{tab-E} lists the final ice thickness for each simulation after the entire heating-cooling event, when the grain has returned to its ambient temperature of 10\,K.

Of all the molecules, the most volatile ones, nitrogen and oxygen, are depleted most. Their evaporation also reduces the grain temperature, making the removal of other species less efficient. For the simulation with highest $T_0=100$\,K, N$_2$ and O$_2$ were completely removed down to the 22nd\,ML, CO to the 18th\,ML, CH$_4$ to the 16th\,ML, with the overall loss of volatiles amounting to 13\,MLs, which is a considerable part of the nonpolar CO-dominated ice layer. Ten percent or more of N$_2$ molecules are removed down to the 30th\,ML, while >10\,\% CO molecules are removed down to the 25th\,ML.

For simulations with lower $T_0$, these molecule removal depths are increasingly lower. Substantial changes in the composition of the surface (e.g., complete removal of a species) occur for $T_0$ down to 40\,K. For $T_0$ values of 50\,K and above, a complete change in the properties of the surface occurs as it changes from nonpolar CO-dominated to polar H$_2$O-dominated.

\section{Conclusions on cosmic-ray induced desorption}
\label{cncl}

%
\begin{table}
\caption{Comparison of molecular CRD yields at $T_0=70$\,K after 10$^{-5}$\,s \citepalias[the approach of][]{Hasegawa93} and the total yield, after complete cooling of the grain from $T_0$ to 10\,K.}
\label{tab-crd}
\centering
\begin{tabular}{c l l}
\hline\hline
 & $Y_{\rm HH93}$ & $Y_{70}$ \\
Molecule & after 10$^{-5}$ s & total \\
\hline
N$_2$ & 4.52E+5 & 1.77E+6 \\
O$_2$ & 8.03E+4 & 3.07E+5 \\
CO & 7.11E+5 & 5.48E+6 \\
CH$_4$ & 2.06E+3 & 5.94E+4 \\
\hline
\end{tabular}
\end{table}
%
   \begin{figure}
		\hspace{-1cm}
    \includegraphics[width=18cm]{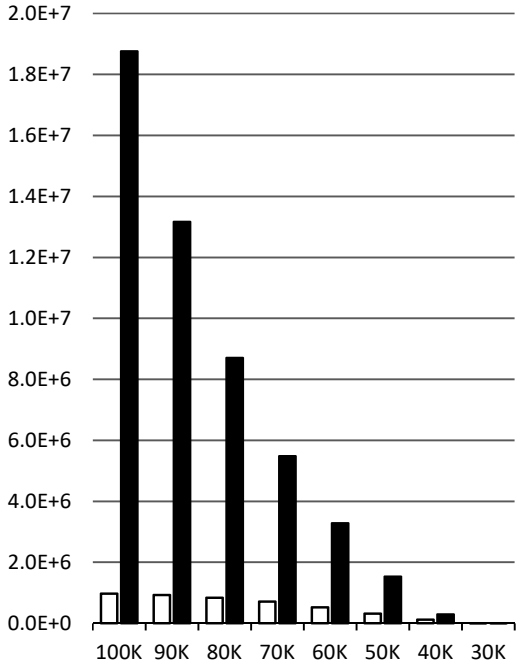}
		\vspace{-15cm}
   \caption{Comparison of evaporation (CRD) yields for the CO molecule from simulations with different $T_0$ values at the time $T_{{\rm evap,CO,}T_0}$ $Y_t$ (empty columns) and total yield $Y$ from the full-length simulations (filled columns). The simulation with $T_0=20$\,K does not produce evaporated CO.}
              \label{att-co}
    \end{figure}
%
   \begin{figure}
		\vspace{-2cm}
		\hspace{-1cm}
    \includegraphics[width=20cm]{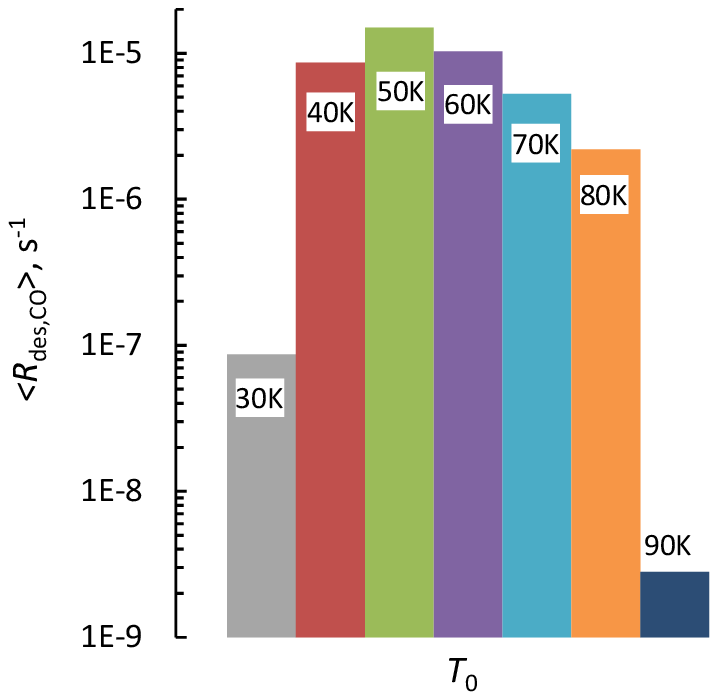}
		\vspace{-20cm}
   \caption{Time-averaged CRD rates of CO molecules for several $T_0$ regimes of CR-induced whole grain heating at an $A_V$ of 10\,mag.}
              \label{att-Rco}
    \end{figure}
The evaporative cooling process has the overall chemical consequence of removing volatile molecules from ice. When the grains have been heated by CRs, the number of molecules that evaporated is the desorption yield $Y_{T_0}$. Figure~\ref{att-co} compares our obtained yields with those obtained by the approach in the original research of \citetalias{Hasegawa93}, $Y_{HH93}$. These authors considered grains heated by iron CR nuclei to 70\,K and assumed that the heated grain cools (and the majority of volatile molecules depart) in a time similar to the evaporation timescale of the CO molecule, which they assumed to be 10$^{-5}$\,s (in our model, the precise $t_{\rm evap, CO}$ is $1.81\times10^{-5}$\,s). A requirement for this approach is that there always must be a sufficient number of molecules available for immediate evaporation. Our study shows that this requirement is not met, primarily because cooling from 70\,K requires removal of several layers of volatiles. Table~\ref{tab-crd} shows that in our model with $T_0=70$\,K, only about one-fourth of N$_2$ and one-eigth of CO molecules have evaporated until 10$^{-5}$\,s, compared to the total yield $Y_{70}$. Grain temperature in the former case drops to only 66.9\,K.

Moreover, the study shows that it is inherently incorrect to relate the CRD yield to the characteristic evaporation timescale of CO (or another coolant species). When it is attributed to different temperature regimes, as was done by \citetalias{KK19}, this approach results in short cooling times at high temperatures and long cooling times at low temperatures. In other words, the assumed timescale is a constant relative to the evaporation rate of the coolant. As a result, the numbers of evaporated molecules are similar for all temperatures, which is obviously incorrect.

Figure~\ref{att-co} demonstrates that the discrepancy of the total CO molecule desorption yield $Y_{70}$ with the yield $Y_t$ resulting from grain cooling until $t_{\rm evap, CO}$ grows with $T_0$. The values of $Y_t$ can be expected to be equal for all $T_0$ values in an ideal case of unlimited evaporation of surface CO. The obtained values are not constant because the model is complex; it considers several coolants, arranged in layers, and radiative cooling. However, for the $T_0$ range of 70-100\,K, $Y_t$ remains fairly unchanged at $(7...10)\times10^{5}$, in line with the discussion above.

These findings make it tempting to reevaluate the results of \citetalias{KK19}. These authors determined that CRD is most effective for grains that are moderately heated by CRs to 40-60\,K. Here, the CRD effectiveness for the important CO molecule (i.e., the average rate of CO desorption, $R_{\rm CRD, CO}$) can be reestimated by multiplying the frequency of the CR-induced grain heating events $f_{T_0}$ \citepalias[from][]{KK19} with the total desorption yield $Y_{T_0}$ for the CO molecule obtained in this work. However, these data are rather incompatible because $Y_{T_0}$ obtained from our study is relevant only for the assumed average ice composition per layer, while $f_{T_0}$ does not have a single value; it is a function of the column density that is traversed by CRs, proportional to the interstellar extinction $A_V$. Nevertheless, we made such an indicative comparison for $A_V=10$\,mag; Figure~\ref{att-Rco} shows that the grain heating to moderate temperatures is still more effective at evaporating CO than low- and high-temperature regimes.

The obtained CRD yields are relevant only to the presented grain model, with its given ice composition, grain size, and ice thickness. These parameters conform to starless cloud core ice that is abundant with volatile molecules. In case of thin ice or ice that is poor in volatiles, the obtained desorption yields overestimate the CRD efficiency for volatile species. On the other hand, the desorption rate of CO$_2$ is likely underestimated for photoprocessed ice that is rich in CO$_2$ and poor in CO. This is because CO$_2$ and other species with similar $E_D$ can become major coolants in the absence of more volatile ice compounds. Moreover, vastly different results can be obtained  when the ice composition is similar but its chemical structure\textup{} is different. We here considered an icy mantle with volatiles that were concentrated mostly in the outer layers. If volatiles lie below tens of layers that are dominated by water, evaporation for them can be expected to be slow and limited.

Massive evaporation at the initial stages of cooling with $T\approx T_0$ depletes the reservoir of available volatile molecules. This reduces the cooling rate and prolongs the time the grain spends in moderate but still elevated temperature. This leaves more time for molecules to overcome barriers in processes other than desorption, such as diffusion or chemical reactions. This can result in an additional chemical processing of icy mantles on grains suffering from CR impacts that heat grains to high initial temperatures \citep{K14,K15aa,Reboussin14}.

\begin{acknowledgements}
JK has been funded by ERDF postdoctoral grant No. 1.1.1.2/VIAA/I/16/194 ‘Chemical effects of cosmic ray induced heating of interstellar dust grains’. The work of JRK has been funded by ERDF project ‘Physical and chemical processes in the interstellar medium’, No 1.1.1.1/16/A/213. Both projects are being implemented in Ventspils University of Applied Sciences. We also thank Ventspils City Council for its persistent support.
\end{acknowledgements}

   \bibliographystyle{aa} 
   \bibliography{dzes1} 

\begin{thebibliography}{35}
\expandafter\ifx\csname natexlab\endcsname\relax\def\natexlab#1{#1}\fi

\bibitem[{{Collings} {et~al.}(2004){Collings}, {Anderson}, {Chen}, {Dever},
  {Viti}, {Williams}, \& {McCoustra}}]{Collings04}
{Collings}, M.~P., {Anderson}, M.~A., {Chen}, R., {et~al.} 2004, \mnras, 354,
  1133

\bibitem[{{Cuppen} {et~al.}(2006){Cuppen}, {Morata}, \& {Herbst}}]{Cuppen06}
{Cuppen}, H.~M., {Morata}, O., \& {Herbst}, E. 2006, MNRAS, 367, 1757

\bibitem[{{Du} {et~al.}(2013){Du}, {Zhao}, \& {Li}}]{Du13}
{Du}, W.-F., {Zhao}, J.-F., \& {Li}, K. 2013, in American Institute of Physics
  Conference Series, Vol. 1547, American Institute of Physics Conference
  Series, ed. L.~{Guo}, 12--20

\bibitem[{{Duley}(1973)}]{Duley73}
{Duley}, W.~W. 1973, \apss, 23, 43

\bibitem[{{Fayolle} {et~al.}(2011){Fayolle}, {{\"O}berg}, {Cuppen}, {Visser},
  \& {Linnartz}}]{Fayolle11}
{Fayolle}, E.~C., {{\"O}berg}, K.~I., {Cuppen}, H.~M., {Visser}, R., \&
  {Linnartz}, H. 2011, \aap, 529, A74

\bibitem[{{Garrod}(2013{\natexlab{a}})}]{Garrod13a}
{Garrod}, R.~T. 2013{\natexlab{a}}, \apj, 765, 60

\bibitem[{{Garrod}(2013{\natexlab{b}})}]{Garrod13b}
{Garrod}, R.~T. 2013{\natexlab{b}}, \apj, 778, 158

\bibitem[{{Hardy}(1947)}]{Hardy47}
{Hardy}, J.~K. 1947, A.R.C reports and memoranda, 2805, 1

\bibitem[{{Hasegawa} \& {Herbst}(1993)}]{Hasegawa93}
{Hasegawa}, T.~I. \& {Herbst}, E. 1993, \mnras, 261, 83

\bibitem[{{Hasegawa} {et~al.}(1992){Hasegawa}, {Herbst}, \&
  {Leung}}]{Hasegawa92}
{Hasegawa}, T.~I., {Herbst}, E., \& {Leung}, C.~M. 1992, \apjs, 82, 167

\bibitem[{{Herbst} \& {Cuppen}(2006)}]{Herbst06}
{Herbst}, E. \& {Cuppen}, H.~M. 2006, Proceedings of the National Academy of
  Science, 103, 12257

\bibitem[{{Iqbal} \& {Wakelam}(2018)}]{Iqbal18}
{Iqbal}, W. \& {Wakelam}, V. 2018, Astronomy and Astrophysics, 615, A20

\bibitem[{{Jenniskens} {et~al.}(1995){Jenniskens}, {Blake}, {Wilson}, \&
  {Pohorille}}]{Jenniskens95}
{Jenniskens}, P., {Blake}, D.~F., {Wilson}, M.~A., \& {Pohorille}, A. 1995,
  \apj, 455, 389

\bibitem[{{Kalv{\= a}ns}(2014)}]{K14}
{Kalv{\= a}ns}, J. 2014, BaltA, 23, 137

\bibitem[{{Kalv{\= a}ns}(2015{\natexlab{a}})}]{K15aa}
{Kalv{\= a}ns}, J. 2015{\natexlab{a}}, \aap, 573, A38

\bibitem[{{Kalv{\= a}ns}(2015{\natexlab{b}})}]{K15apj2}
{Kalv{\= a}ns}, J. 2015{\natexlab{b}}, \apj, 806, 196

\bibitem[{{Kalv{\= a}ns}(2015{\natexlab{c}})}]{K15apj1}
{Kalv{\= a}ns}, J. 2015{\natexlab{c}}, \apj, 803, 52

\bibitem[{{Kalv{\= a}ns}(2016)}]{K16}
{Kalv{\= a}ns}, J. 2016, \apjs, 224, 42 (Paper~I)

\bibitem[{{Kalv{\= a}ns}(2018{\natexlab{a}})}]{K18iau}
{Kalv{\= a}ns}, J. 2018{\natexlab{a}}, in IAU Symposium, Vol. 332, IAU
  Symposium, ed. M.~{Cunningham}, T.~{Millar}, \& Y.~{Aikawa}, 374--380

\bibitem[{{Kalv{\= a}ns}(2018{\natexlab{b}})}]{K18mn}
{Kalv{\= a}ns}, J. 2018{\natexlab{b}}, \mnras, 478, 2753

\bibitem[{{Kalv{\= a}ns} \& {Kalnin}(2019)}]{KK19}
{Kalv{\= a}ns}, J. \& {Kalnin}, J.~R. 2019, \mnras, 486, 2050

\bibitem[{{Kalv{\={a}}ns} {et~al.}(2017){Kalv{\={a}}ns}, {Shmeld}, {Kalnin}, \&
  {Hocuk}}]{K17}
{Kalv{\={a}}ns}, J., {Shmeld}, I., {Kalnin}, J.~R., \& {Hocuk}, S. 2017,
  \mnras, 467, 1763

\bibitem[{{Lacy} {et~al.}(1984){Lacy}, {Baas}, {Allamandola}, {Persson},
  {McGregor}, {Lonsdale}, {Geballe}, \& {van de Bult}}]{Lacy84}
{Lacy}, J.~H., {Baas}, F., {Allamandola}, L.~J., {et~al.} 1984, \apj, 276, 533

\bibitem[{{Leger} {et~al.}(1985){Leger}, {Jura}, \& {Omont}}]{Leger85}
{Leger}, A., {Jura}, M., \& {Omont}, A. 1985, \aap, 144, 147

\bibitem[{{Mart{\'{\i}}n-Dom{\'e}nech}
  {et~al.}(2014){Mart{\'{\i}}n-Dom{\'e}nech}, {Mu{\~n}oz Caro}, {Bueno}, \&
  {Goesmann}}]{Martin14}
{Mart{\'{\i}}n-Dom{\'e}nech}, R., {Mu{\~n}oz Caro}, G.~M., {Bueno}, J., \&
  {Goesmann}, F. 2014, A\&A, 564, A8

\bibitem[{{Merrill} {et~al.}(1976){Merrill}, {Russell}, \&
  {Soifer}}]{Merrill76}
{Merrill}, K.~M., {Russell}, R.~W., \& {Soifer}, B.~T. 1976, \apj, 207, 763

\bibitem[{{Minissale} {et~al.}(2016){Minissale}, {Congiu}, \&
  {Dulieu}}]{Minissale16}
{Minissale}, M., {Congiu}, E., \& {Dulieu}, F. 2016, \aap, 585, A146

\bibitem[{{{\"O}berg} {et~al.}(2009){{\"O}berg}, {Fayolle}, {Cuppen}, {van
  Dishoeck}, \& {Linnartz}}]{Oberg09}
{{\"O}berg}, K.~I., {Fayolle}, E.~C., {Cuppen}, H.~M., {van Dishoeck}, E.~F.,
  \& {Linnartz}, H. 2009, \aap, 505, 183

\bibitem[{{Oort} \& {van de Hulst}(1946)}]{Oort46}
{Oort}, J.~H. \& {van de Hulst}, H.~C. 1946, \bain, 10, 187

\bibitem[{{Reboussin} {et~al.}(2014){Reboussin}, {Wakelam}, {Guilloteau}, \&
  {Hersant}}]{Reboussin14}
{Reboussin}, L., {Wakelam}, V., {Guilloteau}, S., \& {Hersant}, F. 2014,
  \mnras, 440, 3557

\bibitem[{{Tielens} {et~al.}(1991){Tielens}, {Tokunaga}, {Geballe}, \&
  {Baas}}]{Tielens91}
{Tielens}, A.~G.~G.~M., {Tokunaga}, A.~T., {Geballe}, T.~R., \& {Baas}, F.
  1991, \apj, 381, 181

\bibitem[{{van de Hulst}(1946)}]{Hulst46}
{van de Hulst}, H.~C. 1946, Recherches Astronomiques de l'Observatoire
  d'Utrecht, 11, 2.i

\bibitem[{{Watson} \& {Salpeter}(1972)}]{Watson72}
{Watson}, W.~D. \& {Salpeter}, E.~E. 1972, \apj, 174, 321

\bibitem[{{Wickramasinghe}(1965)}]{Wickra65}
{Wickramasinghe}, N.~C. 1965, \mnras, 131, 177

\bibitem[{{Zhao} {et~al.}(2018){Zhao}, {Caselli}, \& {Li}}]{Zhao18}
{Zhao}, B., {Caselli}, P., \& {Li}, Z.-Y. 2018, \mnras, 478, 2723

\end{thebibliography}

\end{document}